\documentclass[preprint, 3p, twocolumn]{elsarticle}

\usepackage{ tabularx }
\usepackage{ graphicx }
\usepackage{ color }
\usepackage{ rotating }
\usepackage{ amssymb }

\newcommand{\ceaual}{CeAuAl$_{3}$}
\newcommand{\ceagal}{CeAgAl$_{3}$}
\newcommand{\cecual}{CeCuAl$_{3}$}
\newcommand{\ceptal}{CePtAl$_{3}$}
\newcommand{\cepdal}{CePdAl$_{3}$}
\newcommand{\cetal}{Ce\textit{T}Al$_{3}$}
\newcommand{\cenial}{CeNiAl$_3$}
\newcommand{\varcenial}{CeNi$_2$Al$_5$}
\newcommand{\ceal}{CeAl$_2$}

\newcommand{\baal}{BaAl$_{4}$}

\newcommand{\cet}{\textit{T}\,=\,Cu, Ag, Au, Pd and Pt}

\begin{document}

\begin{abstract}
We report single crystal growth of the series of \cetal\ compounds with {\cet} by means of optical float zoning. High crystalline quality was confirmed in a thorough characterization process. With the exception of \ceagal, all compounds crystallize in the non-centrosymmetric tetragonal BaNiSn$_{3}$ structure (space group: \textit{I4mm}, No. 107), whereas \ceagal\ adopts the related orthorhombic PbSbO$_2$Cl structure (Cmcm, No. 63). An attempt to grow \cenial\ resulted in the composition \varcenial. Low temperature resistivity measurements down to $\sim$0.1\,K did not reveal evidence suggestive of magnetic order in \ceptal\ and {\cepdal}. In contrast, {\ceaual }, {\cecual} and {\ceagal} display signatures of magnetic transitions at 1.3\,K, 2.1\,K and 3.2\,K, respectively. This is consistent with previous reports of antiferromagnetic order in {\ceaual}, and {\cecual} as well as ferromagnetism in {\ceagal}, respectively.\\
\end{abstract}


\begin{frontmatter}

\title{Single crystal growth of \cetal\ (\textit{T}\,=\,Cu, Ag, Au, Pd and Pt)}

\author[TUM,MLZ]{C. Franz\corref{Cor}}
\ead{christian.franz@frm2.tum.de}

\cortext[Cor]{Corresponding author. +49 89 289 14760  }

\author[MLZ]{A. Senyshyn}

\author[TUM]{A. Regnat}

\author[TUM]{C. Duvinage}

\author[TUM]{R. Sch\"onmann}

\author[TUM]{A. Bauer}

\author[MPI]{Y. Prots}

\author[MPI]{L. Akselrud}

\author[TUM-C]{V. Hlukhyy}

\author[TUM-C]{V. Baran}

\author[TUM]{C. Pfleiderer}

\address[TUM]{Physik Department, Technische Universit\"at M\"unchen, D-85747 Garching, Germany}
\address[MLZ]{Heinz Maier-Leibnitz Zentrum (MLZ), Technische Universit\"at M\"unchen, D-85748 Garching, Germany}
\address[MPI]{Max-Planck Institut f\"ur die chemische Physik fester Stoffe (MPI CPFS), D-01187 Dresden, Germany}
\address[TUM-C]{Fakult\"at f\"ur Chemie, Technische Universit\"at M\"unchen, D-85747 Garching, Germany}

\date{\today}

\end{frontmatter}


\section{Introduction}

Cerium-based intermetallic compounds represent an ideal testing ground for the study of novel electronic ground states and unusual low-lying excitations, where valence fluctuations, heavy-fermion behaviour, unconventional superconductivity and exotic forms of spin and charge order are ubiquitous. Despite many decades of intense research, the understanding of the nature of strong electronic correlations in Ce-based systems, at best, may be referred to as being qualitative. A scenario that is widely alluded to when addressing correlations in f-electron systems considers the competition of Ruderman-Kittel-Kasuya-Yosida (RKKY) interactions, supporting magnetic order, with the single-impurity Kondo effect quenching magnetic moments. While there have been various attempts to advance the understanding, simultaneous treatment of multiple, nearly equivalent energy scales such as exchange and dipolar interaction, spin-orbit coupling, crystal electric fields and strong magneto-elastic coupling has not been attempted.

Given this general context, the class of Ce\textit{T}X$_3$, where T is a transition metal element and X a simple metal, offer important new insights. For instance, selected members of this series have been discovered, which exhibit a coexistence of magnetic order and superconductivity under pressure. Another line of research pursues the strong coupling of phonons with  relatively low lying crystal electric fields (CEF) levels. An important example has been reported in {\cecual} \cite{Adroja2012}, which was interpreted in terms of a quasi-bound vibron state first observed in CeAl$_2$ \cite{loew,thal}. However, recent studies in CePd$_2$Al$_2$, as well as preliminary work in other members of the CeTX$_3$ series suggest, that strong interactions of the crystal fields with the spectrum of phonons somewhat akin the claim of vibrons in {\cecual} is more generic than assumed so far.

In this paper we report single crystal growth of the series {\cetal} with {\cet}. For our study we have used optical float zoning, to the best of our knowledge, for the first time. Following a  thorough characterisation, we determine the crystal structure of the systems studied. The high sample quality achieved in our study is corroborated in measurements of the electrical resistivity, which was performed at temperatures down to 0.1\,K.

\begin{figure}[htbp]
\includegraphics[width=\linewidth,clip=]{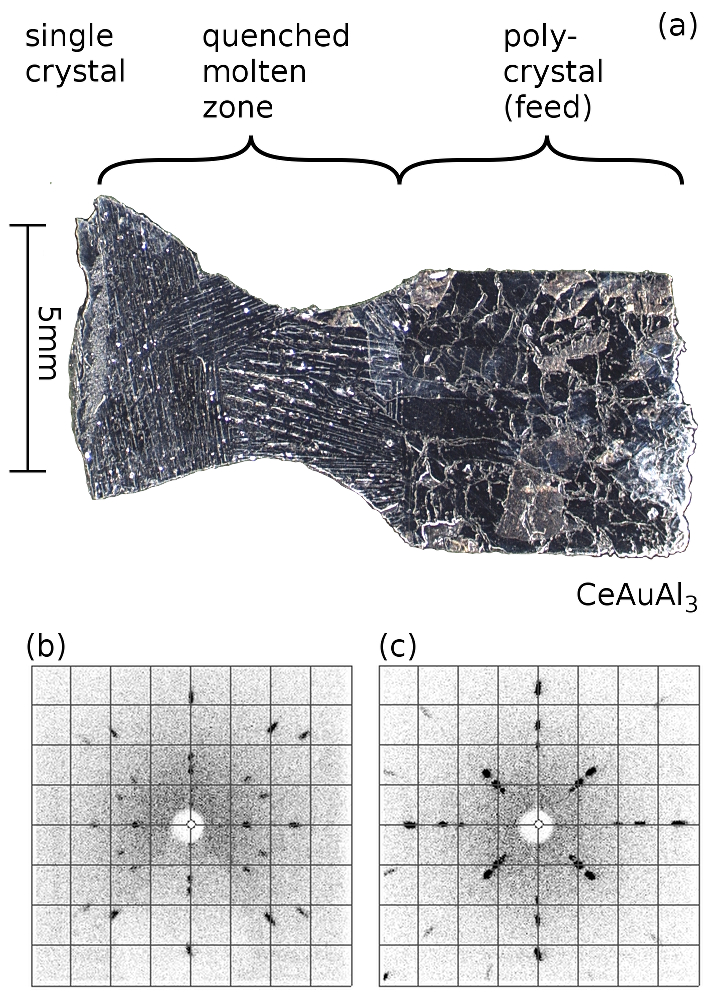}
\caption{(a) Cut through the quenched final zone of the \ceaual\ crystal along the rotational symmetry axis. Polycrystalline part of the feed rod, quenched molten zone and beginning of the single crystal are visible. The quenched molten zone features a pronounced stripe pattern. Laue X-ray images along crystallographic a- and c-axes are shown in panels (b) and (c), respectively. }
\label{pic:zone}
\end{figure}

\begin{figure}[htbp]
\includegraphics[width=\linewidth,clip=]{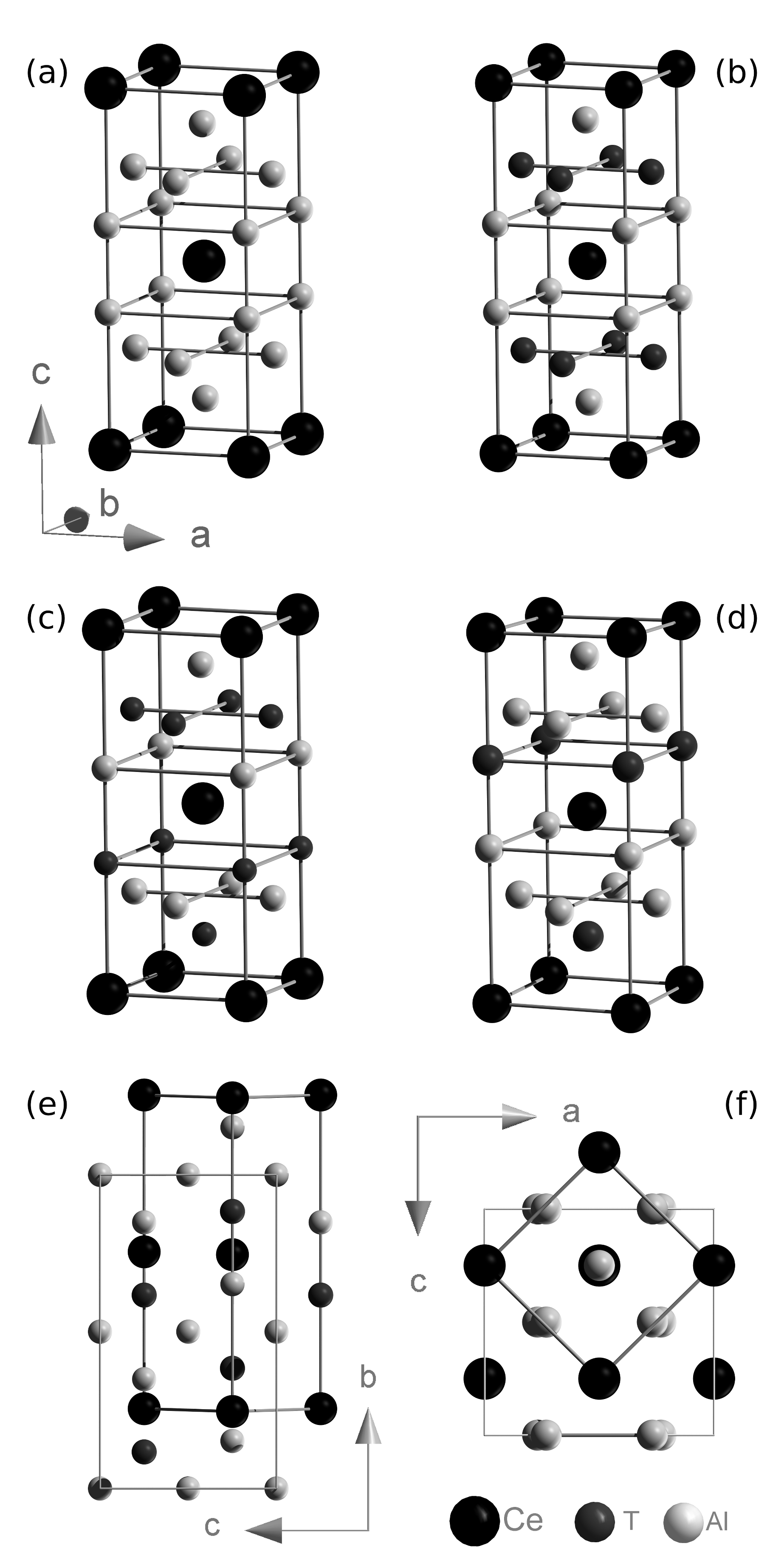}
\caption{Overview over the BaAl$_{4}$ parent structure and important subtypes derived of this structure.
(a) BaAl$_{4}$ with space group \textit{I4/mmm} and full inversion symmetry.
(b) ThCr$_2$Si$_2$ with space group \textit{I4/mmm} and full inversion symmetry.
(c) CaBe$_2$Ge$_2$ with space group \textit{P4/nmm} and full inversion symmetry. 
(d) BaNiSn$_3$ with space group \textit{I4mm}, lacking inversion symmetry.
(e) and (f) Schematic representation of the BaNiSn$_3$ structure type 
and the Cmcm (PbSbO$_2$Cl) structure type.}
\label{pic:strukturen}
\end{figure}

All crystal structures reported in literature for the series of {\cetal} compounds are derived from the BaAl$_4$ structure, depicted in fig. \ref{pic:strukturen}\,(a). The BaAl$_4$ structure crystallizes tetragonal body-centred with eight Ce-atoms at the corners and one in the center. The aluminum atoms are arranged in six planes parallel to \textit{ab} in the unit cell. The following three important types of structures may be derived from the BaAl$_4$ structure in which the Al and \textit{T} atoms are ordered.

(i) The ThCr$_2$Si$_2$ (\textit{I4/mmm}) structure, shown in Fig. \ref{pic:strukturen}\,(b) is currently most frequently reported for intermetallic compounds. The sequence of layers along the \textit{c}-direction is given as A\,-\,[X-T-X]\,-\,A with A a rare-earth metal, T the transition metal element and X either Si, Ge or Al. Typical examples include CeT$_2$X$_2$ compounds (CeCu$_2$Si$_2$ \cite{Edwards1987}, CeCu$_2$Ge$_2$ \cite{deBoer1987}, CePd$_2$Si$_2$, CeRh$_2$Si$_2$ \cite{Ballestracci1978}) as well as URu$_2$Si$_2$ \cite{Cordier1985} and BaFe$_2$As$_2$ \cite{rotter2008}.

(ii) The CaBe$_2$Ge$_2$ (\textit{P4/nmm}) variant of the BaAl$_4$ structure, shown in Fig. \ref{pic:strukturen}\,(c)), is also characterised by full inversion symmetry and a layered structure along the $c$-axis following the sequence A\,-\,[X-T-X]\,-\,A\,-\,[T-X-T]\,-\,A.  The CaBe$_2$Ge$_2$ structure is less frequently found because different atom sizes can not be matched very well in contrast to ThCr$_2$Si$_2$ \cite{Frik2006}. An important example of immediate relevance to the Al-based compounds addressed in our study is CePd$_2$Al$_2$ \cite{pv}.

(iii) The BaNiSn$_3$ (\textit{I4mm}) structure is the only subtype with lacking inversion center. In recent years Ce-systems with a BaNiSn$_3$-type structure have generated great interest, since the discovery of superconductivity in heavy fermion systems such as CeIrSi$_3$, CeRhSi$_3$ \cite{xian1985}, CeCoGe$_3$ \cite{Das2006}. In these systems the superconducting pairing symmetry may be outside traditional classification schemes. Shown in Fig. \ref{pic:strukturen}\,(d) is the characteristic unit cell, where the stacking sequence of layers is A-T-X(1)-X(2)-A-T-X(1)-X(2)-A may be readily seen. The point group of these systems is C$_{4v}$, lacking a mirror plane.

\begin{table*}[htbp]
\begin{center}
\small
\begin{tabularx}{\textwidth}{lp{0.8cm}p{0.95cm}p{2.5cm}p{1.3cm}p{1.2cm}p{1.7cm}p{1.2cm}p{1.7cm}}
		& Mag	& crystal  	& space group							& T$_{\mathrm{C/N}}$			& T$_{\mathrm{K}}$		& $\gamma$ 			& $\mu_\mathrm{eff}$ 	& $\mu_\mathrm{CW}$\\ 
		& 	&		&								& (K)					& (K)				& (mJ/molK$^2$)			& ($\mu_\mathrm{B}$)	& ($\mu_\mathrm{B}$)\\
\hline
\cecual\	& AFM	& pc, sc	& \textit{I4mm} \cite{Moze1996, Mock1999, Klicpera2014cs}	& 2.1 \cite{Bauer1987, Kontani1994}	& 8 \cite{Bauer1987}		& 140 \cite{Bauer1987}		& 1.8 \cite{Mock1999}	& 2.61 \cite{Bauer1987}\\
\ceaual\	& AFM	& pc		& \textit{I4mm} \cite{Hulliger1993, Paschen1998, Sugawara1999}	& 1.32 \cite{Paschen1998}		& 4.5 \cite{Paschen1998}	& 227 \cite{Paschen1998}	& 2.1 \cite{Mock1999}	& 2.6\,-\,2.8 \cite{Sugawara1999}\\
\ceagal\	& FM	& pc		& \textit{I4/mmm} or \newline \textit{I4mm} \cite{Muranaka2007}	& 3.2 \cite{Muranaka2007}		& -				& -				& -			& 2.23 \cite{Muranaka2007}\\ \cepdal\	& AFM	& pc 		& \textit{Fmm2} \cite{Schank1994}		& 6 \cite{Schank1994}					& - 				& - 				& -			& -			\\
\ceptal\	& SG	& pc		& \textit{I4/mmm} \cite{Mock1999, Goerlach2006}			& 0.8 \cite{Goerlach2006, Schank1994}	& -				& -				& 1.8 \cite{Mock1999}	& 2.08 \cite{Goerlach2006}\\
\end{tabularx}
\caption{Resume of \cetal\ properties in literature. AFM - antiferromagnetism, FM - ferromagnetism, SG - Spin-Glass; pc - polycrysal, sc - singlecrystal; T$_{\mathrm{C/N}}$ - Curie/Neel-temperature, T$_K$ - Kondo-temperature; $\mu_\mathrm{eff}$ - effective moment, $\mu_\mathrm{CW}$ - Curie-Weiss moment}
\label{tab:literature}
\end{center}
\end{table*} 

In the following, a brief overview over the literature on \cetal\ compounds is given. Key properties reported are summarised in table\,\ref{tab:literature}. The crystal structure has been determined as \textit{I4mm} in \ceaual\ as well as \cecual\ in both poly- and single crystals. For \ceagal\ no distinction has been possible between \textit{I4mm} and \textit{I4/mmm}, whereas \ceptal\ was reported to adopt a centrosymetric variant \textit{I4/mmm}. Interestingly, \cepdal\ was reported to crystallize in the orthorhombic \textit{Fmm2} structure.

Compared to the textbook examples of 3d magnets, crystal electric fields in 4f compounds are weak. Nevertheless, the crystal fields influence the magnetic properties of 4f systems rather strongly. By Hund's rule, Ce$^{3+}$ has a six-fold degenerate ground state, split into three Kramers doublets by the tetragonal crystal field. In \cecual\ it has been reported that they are low lying with $\Delta_1$\,=15\,K and $\Delta_2$\,=\,238\,K \cite{Bauer1987, Mentink1993}. Further for \ceaual\ excited levels of $\Delta_1$\,=57\,K and $\Delta_2$\,=\,265\,K \cite{Paschen1998} as well as $\Delta_1$\,=60\,K and $\Delta_2$\,=\,240\,K \cite{Sugawara1999} have been reported.

Magnetic ground states in CeTX$_3$ compounds are usually antiferromagnetic, as in \cecual\ with \textit{T}$_N$\,=\,2.1\,K  and \ceaual\ with T$_N$\,=\,1.32\,K. However, the exact magnetic ground states are more complicated, with a propagation vector  \textbf{k}\,=\,(0.4, 0.6, 0) in the case of \cecual\ \cite{Klicpera2015mag} and  \textbf{k}\,=\,(0, 0, 0.52) for \ceaual\ \cite{Adroja2015}. \ceagal\ is a rare example of ferromagnetic Ce-compound below the ordering temperature of T\,=\,3.2\,K. \cepdal\ is reported to order antiferromagnetically below 6\,K, \ceptal\ exhibits spin-glass behaviour below 0.8\,K.

All magnetic compounds posses an easy \textit{ab}-plane and hard \textit{c}-axes. Kondo temperatures have been determined in \cecual\ at 8\,K and \ceaual\ at 4.5\,K, where a weak screening of less than 25\,\% has been found in an $^{27}$Al-NMR study \cite{Vonlanthen1999}. Both compounds show a heavy fermion ground state, demonstrated by an electronic contribution to the specific heat of $\gamma$\,=\,227\,mJ/molK$^2$ as well as a large prefactor of the low-temperature electric resistivity of \textit{A}\,=\,5.0\,$\mu\Omega$cm/K$^2$ in \ceaual\ \cite{Paschen1998} and $\gamma$\,=\,140\,mJ/molK$^2$ in \cecual. Under pressure, T$_N$ rises up to 60\,kbar in \cecual\ and suddenly vanishes at 80\,kbar \cite{Kawamura2010, Nishioka2007}, pointing towards the existence of a quantum critical point (QCP). Recently, a phonon - crystal field quasibound state has been found in \cecual\ \cite{Adroja2012}, which is not present in \ceaual\ \cite{Adroja2015}. Furthermore, a low energy anomalous excitation \cite{Aoki2000a} and phonon scattering by Ce magnetic moment fluctuations are reported in \ceaual\ \cite{Aoki2000}, as well as a possible second phase transition at 0.18\,K \cite{Adroja2015}.

\section{Experimental Methods}

The physical properties of rare earth containing compounds tend to be very sensitive to defects and impurities. In turn, we have made great efforts to reduce such defects and impurities to the lowest possible level. Notably, the procedure of the single-crystal preparation is based on the use of high purity starting elements and a bespoke work-flow that is optimised to minimise contamination by oxygen. As the perhaps most important aspect, we have used optical float-zoning (OFZ) representing a crucible-free technique.

\begin{figure}[htbp]
\begin{center}
\includegraphics[width=0.8\linewidth]{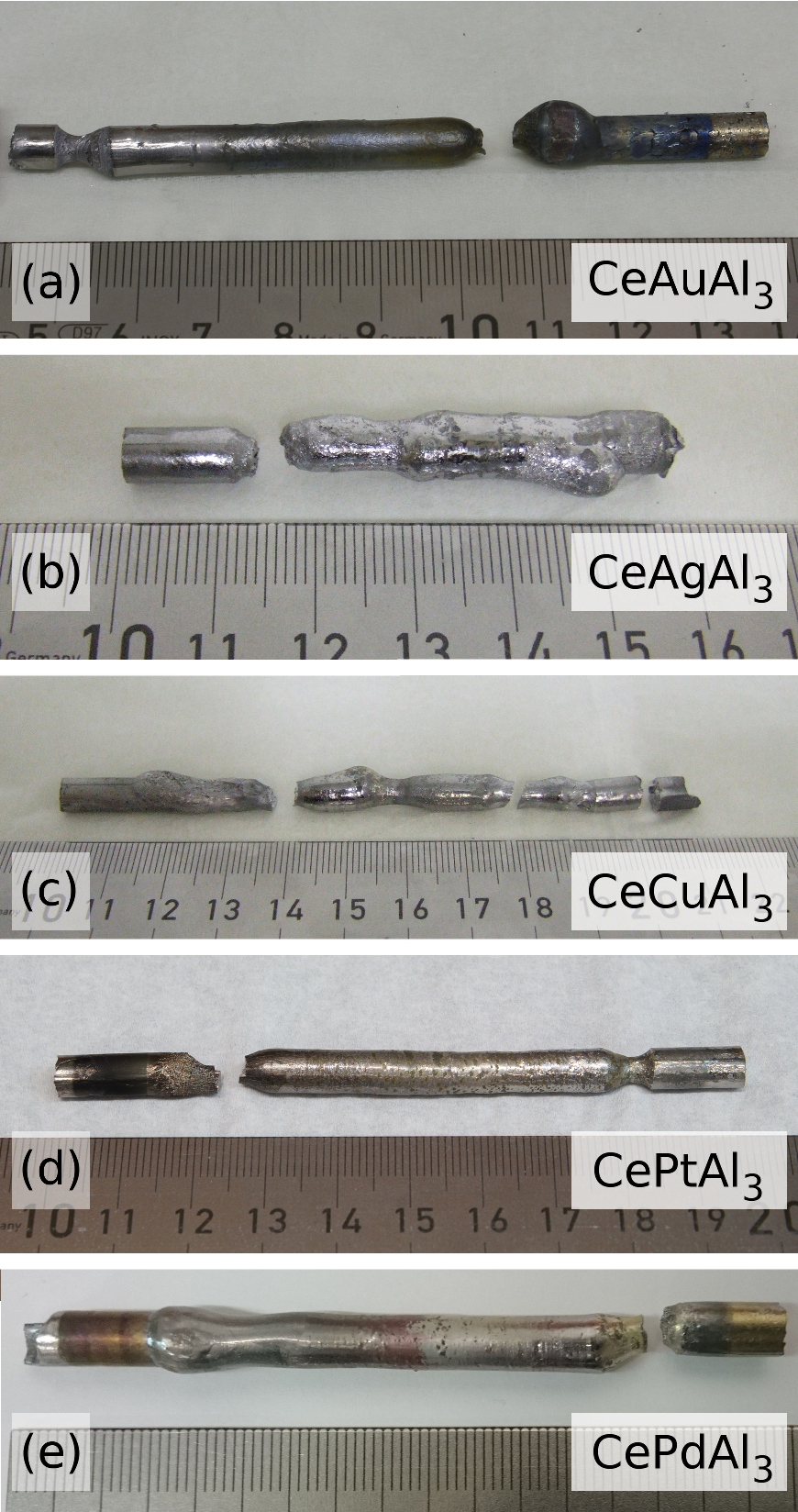}
\caption{Photographs of all crystals grown by optical float zoning for this study. Scale in cm.
         (a) \ceaual\
         (b) \ceagal\
         (c) \cecual\
         (d) \ceptal\
         (e) \cepdal}
\label{pic:photo}
\end{center}
\end{figure}

Metal lumps of Ce (Ames, 99.9995\%), shots of Au and Ag (AlfaAesar Premion, 99.995\% 99.9999\%), lumps of copper (MaTecK, 99.9999\%), 
powder of Pd and Pt (AlfaAesar Premion, both 99.995\%) and lumps of Al (MaTecK, 99.9999\%) were used for the preparation of the feed and seed rods. Ce and Al were weighted in an argon glove box system. First \ceal\ was prepared in an induction heated horizontal copper cold boat system that can be loaded by means of a bespoke load-lock from a glove box  \cite{Bauer2016}. This reduces contamination of the Ce with oxygen, as the educt \ceal\ is stable on air. In a second step, the \ceal\ was then reacted with the transition metals \textit{T}\,=\,Au,\,Ag,\,Cu,\,Pt or Pd and additional Al in a water cooled Hukin-type radio frequency heated copper crucible. 
 
The sample was remelted and flipped over several times to assure a good homogeneity. The resulting pill was subsequently cast into cylindrical feed and seed rods with a diameter of  6\,mm and a length of 40\, to \,60\,mm's. Furthermore, synthesis of a polycrystalline sample of \cenial\ was attempted by means of our rod casting furnace. For \ceaual\ the ternary compound was prepared without the intermediate step of first synthesising {\ceal} by means of the glove box and the horizontal cold boat system and the bespoke load-lock.
 
The actual single crystal growth was performed in a CSI 4-mirror (model FZ-T 10000-H-III-VPS) optical float zone furnace. The furnace was redesigned to be all-metal sealed. Prior to growth the furnace was baked with bespoke heating jackets \cite{Neubauer2011}. 500\,W halogen lamps were used for the growth process under a static argon atmosphere at a pressure of 2.5\,bar. The float zoning was performed with a growth rate of 5\,mm/h  and a counter rotation of 6\,rpm of the feed and seed rod, respectively.

The characterization of the crystals proceeded as follows. First, all crystals were examined under an optical light microscope. Laue X-ray (Multiwire Laboratories MWL 110) images were taken on different spots covering all of the surface of the ingot to identify the single crystalline part and possible grain boundaries. In \ceaual\ additionally cross sections of the upper and lower part of the crystal were analysed as well as a cut through the quenched final part of the molten zone. Figure \ref{pic:zone} (a) shows from right to left the polycrystalline feed rod, the quenched molten zone with a pronounced stripe structure and the beginning of the single crystal at the lower end.

All crystals were oriented by Laue X-ray diffraction. Fig. \ref{pic:zone}\,(b) and (c) show typical Laue pictures of \ceaual\ for the \textit{a}- and \textit{c}-axes, respectively. The composition was confirmed by single crystal and powder x-ray diffraction. For \ceaual\ additional scanning electron microscope images and energy dispersive X-ray spectra were recorded.

Small single crystals with dimensions less than 100\,$\mu$m were mechanically extracted from  \cetal\ (\cet) ingots as grown. Generally all tested crystals were of high quality with sharp diffraction peaks. Intensity data were collected with graphite-monochromatized Mo K$\alpha$ X-ray radiation. Three-dimensional data were indexed, integrated and corrected for Lorentz-, polarization, absorption and background effects using a diffractometer specific software, namely CrystalClear for a Rigaku Saturn724+ and X-Area for a STOE IPDS II. Initial structure analysis/solution (using direct methods) for \cetal\ (\textit{T}\,=\,Ag, Au, Cu, Pt) was done with SHELX-2012 as implemented in the program suite WinGX 2014.1 \cite{Farrugia2012}. The crystal structure of modulated \cepdal\ was solved using WinCSD \cite{Akselrud2014}.The experimental data and results of the structure refinement for selected samples are reported in table \ref{tab:PowderDiff}, while the fractional atomic coordinates, occupation numbers and equivalent isotropic, and anisotropic atomic displacement parameters are listed in table \ref{tab:S1} and \ref{tab:S2}.

Samples for resistivity measurements were cut from the single-crystalline sections of the ingots with a diamond wire saw. Typical dimension of the samples were 4\, to 6\, times 1\, times 0.2\,mm$^3$, oriented such that the electrical current could be applied along longest direction which corresponding to the \textit{c}-axis. The resistivity was measured in a standard four-probe ac-configuration using a digital lock-in amplifier and an excitation current of 5\,mA at an excitation frequency of 22.08\,Hz. Room temperature transformers were used for impedance matching. Data were recorded between 2\,K and 300\,K in a standard $^4$He cryostat. Data down to much lower temperatures were measured for \cetal\ with \textit{T}\,=\,Cu, Ag, Au in an adiabatic demagnetisation refrigerator (ADR) using the same detection technique at a lower excitation current of 1\,mA down to a temperature below $\sim$300\,mK. For samples with \textit{T}\,=\,Pt, Pd a $^3$He/$^4$He dilution refrigerator was used down to temperatures well below 100\,mK.


\section{Experimental Results}

\subsection{Crystal Structure}

The actual growth process resulted in the ingots shown in Fig.\,\ref{pic:photo}. We obtained large single crystals of \ceaual, \cecual\ and \ceptal\ shown in Figs.\,\ref{pic:photo}\,(a), (c) and (d), respectively. While these ingots remained mechanically intact after growth, the \cecual\ ingot (shown in Fig.\,\ref{pic:photo}\,(a)) broke spontaneously in one location during cool-down after growth. For \ceagal\ single crystalline grains were prepared from the ingot. The melt in the growth process of \cepdal\ was very unstable and we could only prepare small single crystalline samples with a typical size of up to 3\,mm. A second attempt to grow {\cepdal} with a reduced growth rate of 1\,mm/h allowed to obtain the large single crystal depicted in Fig.\,\ref{pic:photo}\,(e). Note that all measurements on {\cepdal} reported in this paper were carried out on samples cut from the first crystal.


The analysis of the arrays of the diffraction data revealed that the majority of the \cetal\ compounds (\textit{T}\,=\,Au, Cu, Pt) studied display reflections consistent with the tetragonal lattice and in line with one of distorted \baal\ structure types. The T/Al antisite disorder essentially corresponds to the local differences between ThCr$_2$Si$_2$, CaBe$_2$Ge$_2$, HoCuAl$_3$ or BaNiSn$_3$ structures. Extinction of characteristic reflections revealed a body centred tetragonal lattice and the structure solution corresponded to a BaNiSn$_3$ type structure for \ceaual, \cecual\ and \ceptal. Evidence for putative Au/Al antisite disorder in \ceaual\ was below the detection limit, whereas for \cecual\ a small Cu/Al mixing on the copper site was observed consistent with a $\sim$5\,\% site occupation factor (sof) deficiency on the 4b Al site. A higher degree of antisite disorder was found in our \ceptal\ sample. Assuming a stoichiometric composition and fully occupied atomic sites of \ceptal, the antisite Pt/Al disorder appears to be as high as 18\,\% sof at both T and Al 2a sites. The obtained unit cell volumes  were found to decrease in the series Au -- Pt -- Cu in accordance with the metallic radii (r$_{Au}$\,=\,1.44\,\AA, r$_{Pt}$\,=\,1.39\,\AA, r$_{Cu}$\,=\,1.28\,\AA).

For \ceagal\ a small orthorhombic distortion of the BaAl$_4$ type lattice was detected of order $\sim$0.1\,\AA. The character of the reflection splitting and the extinction revealed a C-base centred orthorhombic lattice with a$_0\approx$\,a$_T\sqrt{2}$, b$_0$\,$\approx$\,c$_T$ and c$_0$\,$\approx$\,a$_T\sqrt{2}$, where “O” and “T” correspond to orthorhombic and tetragonal lattice dimensions. The structure solution corresponds to the \textit{Cmcm} space group and a model consistent with the structure type of PbSbO$_2$Cl. The orthorhombic superstructure observed occurs in oxyhalides and is found rarely in intermetallics, e.g., in LaZn$_4$ \cite{Oshchapovsky2012} and SrPdGa$_3$ \cite{Seidel2014}. Similar to \cecual\ a small Ag/Al mixing occurs on the Ag site along with a weak (ca. 4\,\% sof) deficiency on the 8e Al site. 

A comparison of the Ag and Au based \cetal\ structures is plotted in Fig.\,\ref{pic:strukturen}\,(e) and (f). We note, that Ag and Au nominally possess the same metallic radii (r$_{Ag}$=1.44\,\AA), which may correspond to a similar magnitude of chemical pressure in both \cetal\ (\textit{T}\,=\,Au, Ag). Indeed, the normalised lattice dimensions of \ceagal\ (a$_T$\,$\approx$\,(a$_0$+b$_0$)/2$\sqrt{2}$\,$\approx$\,4.3566(8)\,\AA, c$_T$\,$\approx$\,10.837(2)\,\AA) are very similar to those of \ceaual. Further, the relative cerium and aluminium positions reproduce well, whereas major differences were noticed for the distribution of gold and silver sites. By their arrangement the structure may be viewed as an intermediate step between BaAl$_4$ (or HoCuAl$_3$) and the BaNiSn$_3$ structure type. The antisite disorder observed smears out the layered structure along the c-axis in \ceagal, which initially can be described as A - [T{X}-X-T{X}] – A, where T{X} indicated mixed T and X layer occupation.

\begin{figure}[htbp]
\begin{center}
\includegraphics[width=0.9\linewidth]{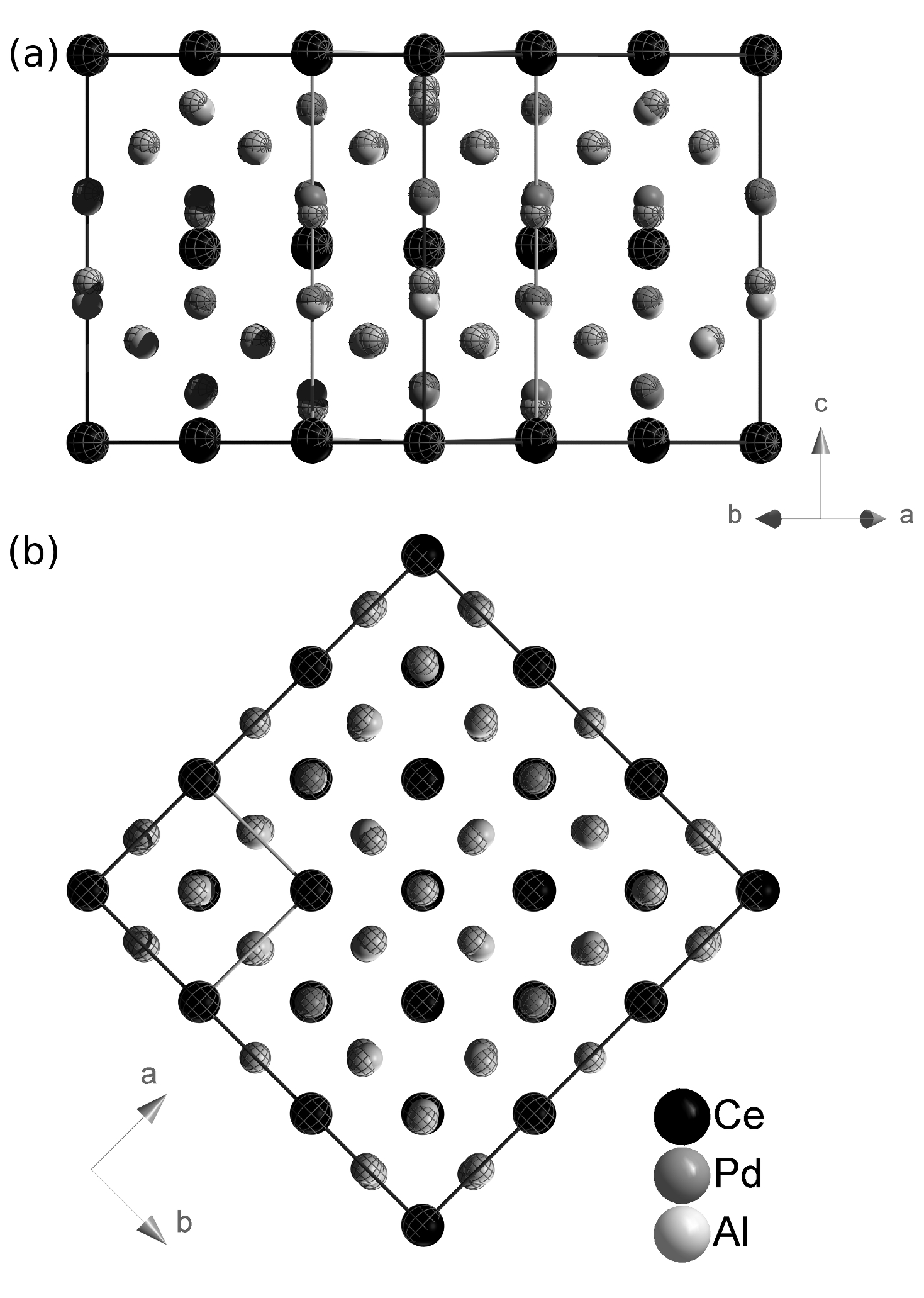}
\caption{Refined structural modifications of \cepdal: plane atoms - BaNiSn$_3$ type of structure, structured atoms - klassengleiche subgroup of BaNiSn$_3$ type of structure with 3\textit{a} multiplied axis.}
\label{pic:anat_2}
\end{center}
\end{figure}

\begin{figure}[htbp]
\includegraphics[width=\linewidth]{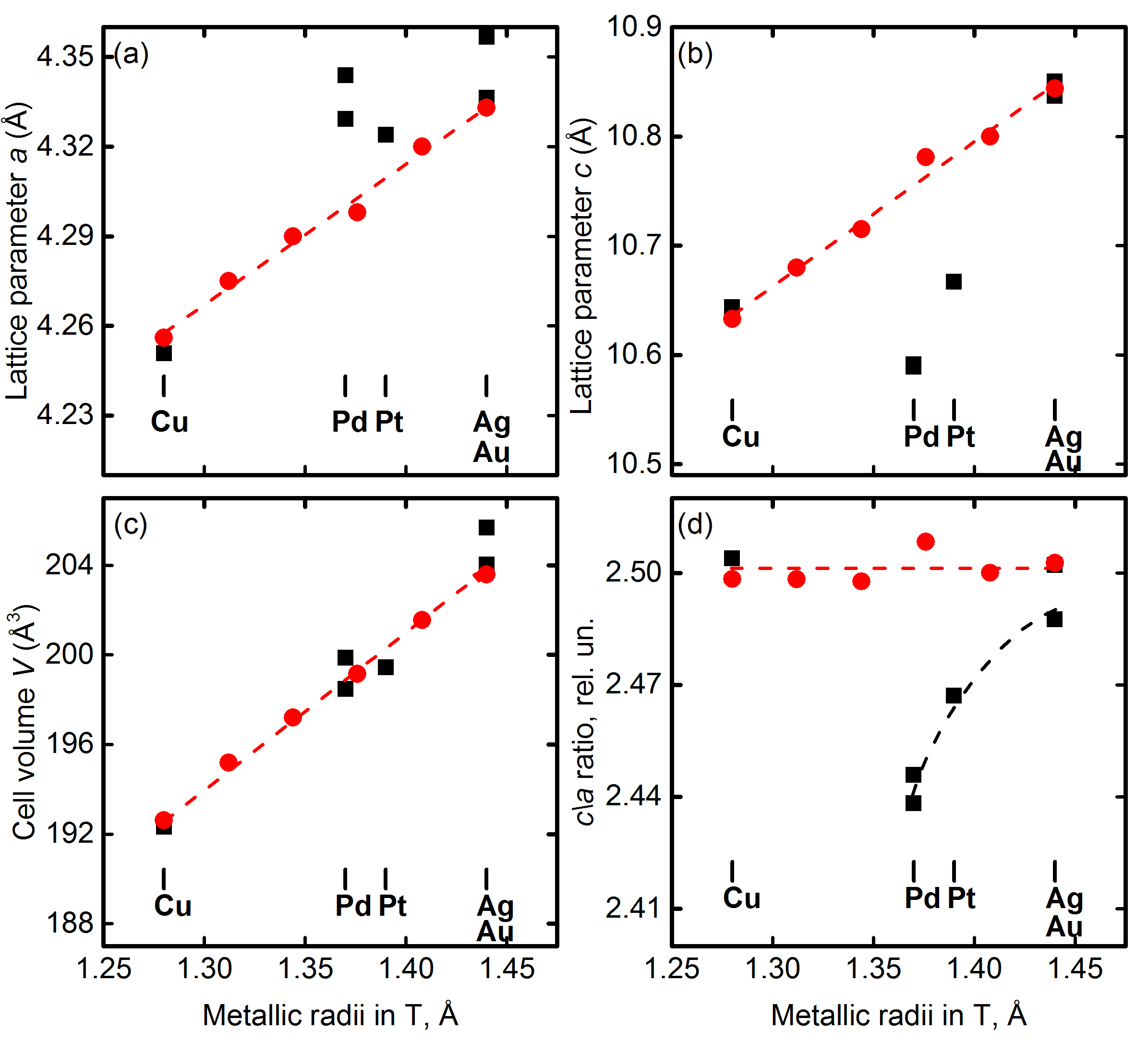}
\caption{Lattice parameter, cell volume and c/a ratio for different \cetal\ (\textit{T}\,=\,Ag, Au, Cu, Pd, Pt) plotted as a function of T metallic radius (black squares). Data showed by red circles correspond to CeAu$_{1-x}$Cu$_x$Al$_3$ solid solution taken from \cite{Klicpera2014}. Two data points for \cepdal\ correspond to BaNiSn$_3$ and “3a” structures. The lines are showed as guides for the eyes.}
\label{pic:latticeParams}
\end{figure}

The set of Bragg reflections collected for our single crystal of \cepdal\ could neither be described with a conventional tetragonal nor with orthorhombic lattices. Careful indexing reveals a tetragonal cell with a \textit{c}-parameter comparable to \cetal\ (\textit{T}\,=\,Au, Ag, Cu, Pt), however, with an \textit{a} lattice parameter multiplied by a factor of $\sim$\,3. Analysis of peak systematics resulted in a solution corresponding to a body centred tetragonal lattice. This is in agreement with group theory, as the 3\textit{a}, \textit{c} axis multiplication in the \textit{I4mm} space group is allowed in the frame of klassengleiche subgroups IIc for \textit{I4mm} (“3a” structure). Multiplication of the axis results in a splitting of the Ce and Pd positions into three independent sites, whilst the two initial aluminium sites are split into seven positions in the larger cell. Similar to \cecual, \ceptal, \ceagal\ Rietveld refinement corresponded to a Pd/Al antisite disorder, where, however, only one Pd site of three as well as two of the seven Al sites were affected.

Examination of a crushed and pulverized ingot using lab X-ray powder diffraction did not reveal any hints for a 3\textit{a} structure, but a conventional BaNiSn$_3$ modification for \cepdal\ (see Table \ref{tab:S2}). The crystal structures of both the BaNiSn$_3$-type and the 3\textit{a}, \textit{c} axis multiplied modifications of the \cepdal\ structures are shown in Fig.\,\ref{pic:anat_2}. The Ce(0,0,0) atomic position is here used as a reference. In comparison to the BaNiSn$_3$-type modification of \cepdal, the 3\textit{a} structure contains a certain degree of Pd/Al disorder and by its layer structure A - [T{X}-X-T{X}] - A it becomes very similar to \ceagal. An attempt to refine the structure modulation in \cepdal\ was performed assuming both commensurate and incommensurate modulation vectors of the BaNiSn$_3$ parent structure. The atomic modulated displacements were comparable to standard uncertainties of atom localization and improvement to fit residuals (when compared to the structure model with 3\textit{a}, \textit{c} axis multiplication), and thus, found to be marginal.

The lattice parameters of the \cetal\ (\textit{T}\,=\,Ag, Au, Cu, Pd, Pt) samples studied were found to be in a good agreement with experimental data reported in Refs.\,\cite{Klicpera2014, Moze1996, Zare1965, Mentink1993, Kontani1994}. Cell volumes in \cetal\ (\textit{T}\,=\,Ag, Au, Cu, Pd, Pt) follow a linear dependence as a function of T ionic radii (see Fig. \ref{pic:latticeParams}) in line with values reported for the \ceaual\ to \cecual\ pseudobinary system \cite{Klicpera2014}, which displays a behaviour characteristic of Vegards law. The observed linear dependence may be viewed in terms of structural changes in \cetal\ (\textit{T}\,=\,Ag, Au, Cu, Pd, Pt) driven by “chemical pressure”. However, all \cetal\ studied are characterised by a large anisotropy. Notably, for \cecual, \ceaual\ and corresponding solid solutions \cite{Klicpera2014} the c/a ratio is nearly constant with c/a\,$\sim$\,2.5. This compares with \cepdal\, \ceptal\ and {\ceagal}, which display lower c/a values. 

Taking a separate look on the experimentally determined lattice parameters indicates \textit{a} parameters that are systematically reduced, as well as \textit{c} lattice parameters that are systematically smaller than those observed in the \ceaual\ -- \cecual\ pseudobinary system. Taking into account nominally equal metallic radii for Au and Ag, the differences observed for the a lattice dimensions and cell volume of \ceaual\ and \ceagal\ as well as their different structure, significantly reduces the relevance of an approximation in the spirit in terms of  “chemical pressure” and shows that the chemical nature of T in \cetal\ plays a dominant role for the structure and its distortion.

\subsection{Electric Resistivity}

\begin{figure*}[htbp]
\includegraphics[width=\linewidth]{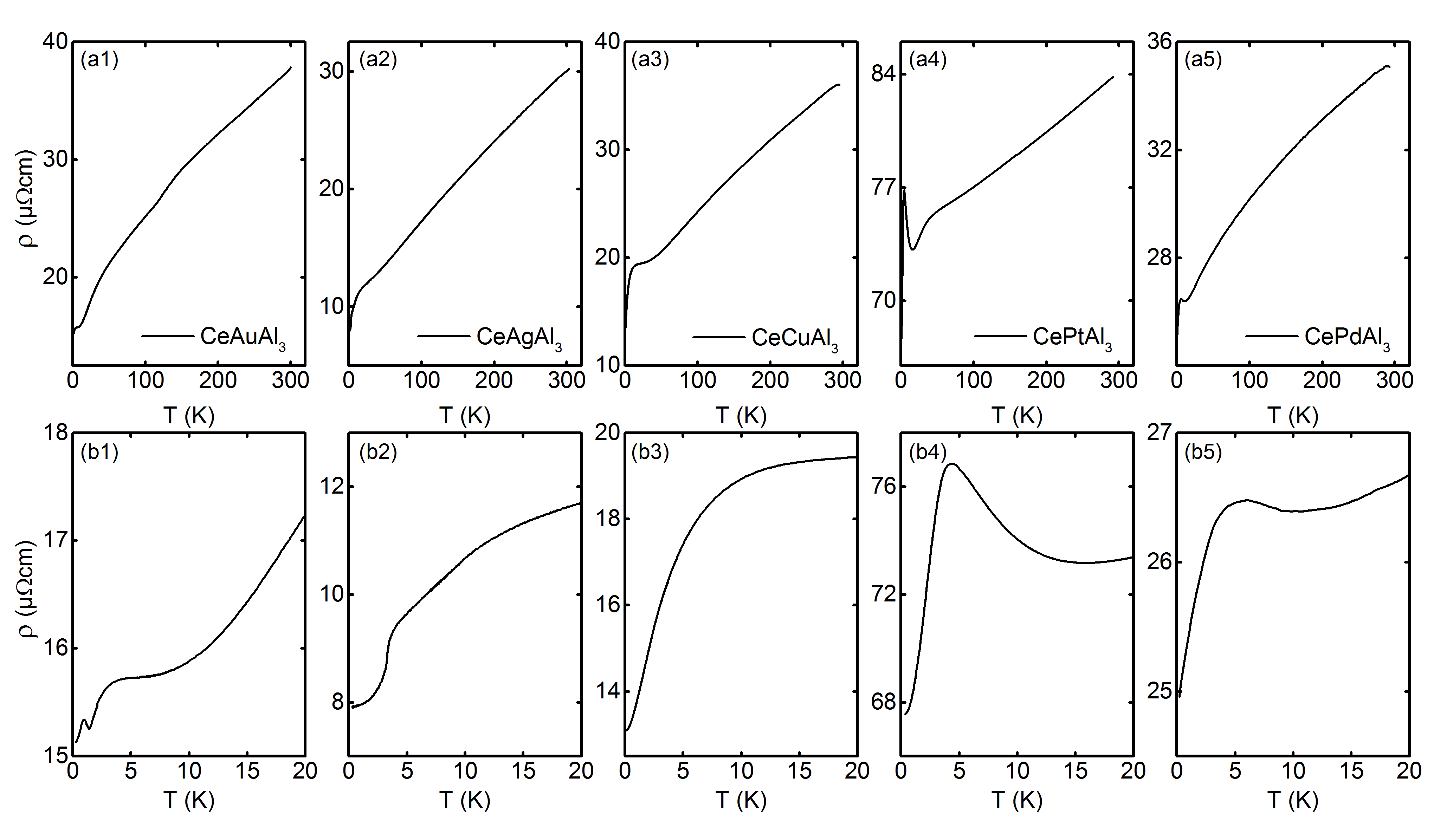}
\caption{Electric resistivity for \cetal\ with \textit{T} = Cu, Au, Ag, Pd, Pt. (a1) - (a5) On a large temperature range from T$_{min}$\,=\,300\,mK to 300\,K all samples show metallic behaviour.  (b1) - (b2). Close-up view of the behaviour at low temperatures up to 20\,K.}
\label{pic:resistivity}
\end{figure*}

\begin{table}[tb]
\begin{center}
\small
\begin{tabular}{lccc}
		& $\rho_0$ ($\mu\Omega$cm)	& RRR	& $\rho_{300\,K}$ ($\mu\Omega$cm)	\\
\hline
\ceaual\  	& 15.18 			& 2.50 	& 37.8   	\\
\cecual\  	& 12.9 				& 2.84 	& 36.6   	\\
\ceagal\  	& 7.9 				& 3.81 	& 30.1   	\\
\cepdal\  	& 25.1 				& 1.39 	& 35.0	   	\\
\ceptal\  	& 67.4 				& 1.25 	& 84.1      	\\
\end{tabular}
\caption{Parameters derived from resistivity measurements on \cetal\ with \textit{T}\,=\,Au,Cu,Ag,Pd and Pt. $\rho_0$ is the residual resistivity extrapolated to T\,=\,0, RRR the residual resistivity ratio $\rho$\,(300\,K)/$\rho_0$, where $\rho$\,(300\,K) denotes the resistivity at room temperature.}
\label{tab:res}
\end{center}
\end{table}

Shown in Fig.\,\ref{pic:resistivity}\,(a1) to (a5) is the electric resistivity for \cetal\ with \cet\ covering three decades of temperature from $\sim$0.1\,K to 300\,K. For these data the electrical currents were applied parallel to the crystallographic \textit{c}-axis. Further, Fig. \ref{pic:resistivity} (b1) - (b5) shows the same data on a temperature scale up to 20\,K for a better visibility of the low temperature features. 

All samples show metallic behaviour over the complete temperature range. The residual resistivity, resistivity at 300\,K and associated residual resistivity ratios (RRR) of all compounds are summarized in table \ref{tab:res}. The residual resistivity ratios below 5 for all compounds and are in good agreement with the literature. Observed values have been found lower than in Si or Ge based compounds, probably due to the remaining antisite disorder in our samples. We wish to note that preliminary work suggests, that careful annealing studies and tiny compositional adjustments to the starting compositions as well as variations of the parameters used for single crystal growth promise considerable improvements to these values. 

The key features observed in the temperature dependences may be summarised as follows. For \ceaual\ the resistivity decreases of from 300\,K to 15\,K in a quasi-linear manner, probably due to crystal electric field levels as discussed in ref. \cite{Paschen1998}. With decreasing temperature the decrease of the resistivity is followed by a plateau around 8\,K and another strong decrease below the onset of magnetic order at 1.32\,K. However the onset of the decrease at 4.5\,K is considerably higher than $T_c$ and maybe attributed to the coherence of a Kondo lattice. At $\sim$1\,K, an additional maximum due to the opening of a super-zone gap is observed reminiscent of behaviour observed in CeGe \cite{Das2012} and Cr \cite{Stebler1970}.

A linear decrease is observed in \ceagal\ down to approximately 17\,K, followed by a steeper linear decrease. The change of slope is of unknown origin, but may be due to the changes of the coupling with the lattice degrees of freedom. A clear kink at 3.2\,K, corresponding to the magnetic transition temperature of ferromagnetic order reported in the literature, is followed by a quadratic temperature dependence down to lowest temperatures. The specific temperature dependence may be characteristic of weakly spin-polarised Fermi liquid, where electron magnon scattering is weak.

In {\cecual} the linear decrease at high temperatures is followed by a plateau between 30\,K and 10\,K, whereas a clear maximum as reported in Refs\,\cite{Kontani1994, Klicpera2015mag} is not observed. The strong decrease below 10\,K traditionally would be attributed to the coherence of a Kondo lattice, where a clear signature of the magnetic ordering transition at 2.1\,K is not observed and maybe hidden by this Kondo coherence. However, on a speculative note an alternative scenario may be related to the strong electron-phonon interactions associated with the quasi-bound vibron. Below 0.5\,K the decrease flattens towards a residual resistivity of 12.9\,$\mu\Omega$cm, which is among the lowest reported in the literature for this compound.

The linear decrease of the resistivity in \ceptal\ down to $\sim$\,50\,K is followed by a stronger decrease to a minimum at 15.9\,K. The origin of this kink may again be referred to as a Kondo coherence effect and yet, as speculated above, related to changes of the coupling to the crystal lattice.  A logarithmic increase to a maximum at 4.3\,K is most probably due the onset of Kondo scattering and followed by a strong decrease suggesting again Kondo coherence. Towards lowest temperatures, the decrease flattens and  is extrapolated to a residual resistivity of 25.1\,$\mu\Omega$cm. Overall, the resistivity is about a factor of three higher than for other compounds studied, which may be due to the higher degree of antisite disorder. There are no signs suggesting magnetic order, however below $\sim$0.8\,K the decrease of the resistivity flattens out.

Last but not least, \cepdal\ shows a sublinear decrease over the entire temperature range, probably due to crystal field effects. The overall behaviour with a faint maximum around 5.7\,K and strong decrease down to the lowest temperatures very much resembles the behaviour of \ceptal. However, both the maximum and the decrease to lowest temperatures are less pronounced. Towards the lowest temperatures there is a linear decrease of the resistivity instead of a flattening out. As for \ceptal\ we do not observe any features suggesting magnetic order.

\section{Conclusions}

Single crystals of the non-centrosymmetric intermetallic compounds \cetal\ (\cet) were grown. All samples were grown by means of optical float-zoning, to the best of our knowledge for the first time. Large single crystals were obtained for all compounds except \ceagal. For \cepdal\ the growth rate had to be reduced from 5\,mm/h to 1\,mm/h to be stable. An attempt to grow a poly-crystal of \cenial\ resulted in the compound \varcenial. The crystal structure was determined by single crystal and powder X-ray diffraction. The crystal structure is \textit{I4mm} in all cases except for \ceagal, which shows a small distortion corresponding to the orthorhombic \textit{Cmcm} structure. Further, the space group of \cepdal\ has a three fold multiplied lattice parameter a, which is allowed by group theory. Although the unit cell volume follows the metallic radii of the transition metal element rather well, the c/a ratio is deviating strongly.

Site-antisite disorder of the T and X-site appears to be nominally absent in {\ceaual}, but reaches a value as high as 18\,\% in \ceptal, accounting for the larger residual resistivities as compared to the Si and Ge based compounds in this series. While the sample quality of our samples as grown is already high, we expect that careful studies addressing the precise growth conditions and role of post-growth treatment, such as annealing and/or electro-transport, promise significant improvements. 

Further, resistivity measurements as described in the traditional language used for Ce-based compounds are consistent with Kondo behaviour for all samples except \ceagal. The occurrence of the Kondo maximum depends thereby on the transition metal element and is most pronounced for \textit{T}\,=\,Pt. It is interesting to speculate, if, what seems to be Kondo-like behaviour, eventually turns out to be due to strong electron-phonon coupling as related to the notion of the quasi-bound vibron state reported in {\cecual}. As a final point, features in the resistivity of {\ceaual}, {\cecual} and {\ceagal} at low temperatures are consistent with published reports of antiferromagnetic order in the former two compounds and ferromagnetism in the latter system, respectively. This contrasts the literature on \ceptal\ and {\cepdal}, where we do not find signs of magnetic order down to 0.1\,K in the resistivity, whereas spin glass order has been reported for \ceptal\ and antiferromagnetic order for \cepdal.

Taken together all single-crystals are of high quality consistent with very low residual resistivities as compared to the literature. This demonstrates that optical float-zoning as employed under pure conditions provides an excellent method for the preparation of Ce-based intermetallic compounds.

\section{Acknowledgements}
We wish to thank Matthias Ruderer for access to the glove box of E13,
Rainer Jungwirth (FRMII) for EDX measurements on \ceaual, the TUM crystal lab for crystal orientation with Laue X-ray and preparation of the samples. Financial support of the Deutsche Forschungsgemeinsschaft and DFG TRR80 are gratefully acknowledged.

\section*{References}
\bibliographystyle{plain}	
\bibliography{citations}	

\clearpage

\begin{table*}[htbp]
\begin{center}
\small
\begin{tabularx}{\textwidth}{p{2.5cm}lllll}
		& \ceagal\ 	& \ceaual\		& \cecual\	& \cepdal\	& \ceptal\ \\
\hline
Diffractometer	& Stoe IPDS-II	& Rigaku Saturn724+	& Stoe IPDS-II	& Stoe IPDS-II 	& Rigaku Saturn724+ \\
Radiation	& MoK$\alpha$	& MoK$\alpha$		& MoK$\alpha$	& MoK$\alpha$	& MoK$\alpha$ \\
Wavelength (\AA)& 0.71073	& 0.71073		& 0.71073	& 0.71073	& 0.71073 \\
Abs. correction type & empirical & empirical 		& empirical 	& empirical 	& empirical \\
Abs. coeff. $\mu$ & 15.737	& 237.31		& 16.792	& 161.11	& 232.91 \\
Crystal system	& orthorhombic	& tetragonal		& tetragonal	& tetragonal	& tetragonal \\
space group	& Cmcm		& I4mm			& I4mm		& I4mm		& I4mm \\
a (\AA)		& 6.2050(12)	& 4.3364(4)		& 4.2508(4)	& 12.988(1)	& 4.3239(4) \\
b (\AA)		& 10.837(2)	& 4.3364(4)		& 4.2508(4)	& 12.988(1)	& 4.3239(4) \\
c (\AA)		& 6.1176(12)	& 10.8501(15)		& 10.6436(13)	& 10.589(1)	& 10.6670(15) \\
V (\AA$^3$)	& 411.38(14)	& 204.03(4)		& 192.32 (3) 	& 1786.2(5)	& 199.43(4) \\
Z		& 4		& 2			& 2		&		& 2 \\
$\rho_\mathrm{calc}$ (g/cm$^3$) & 5.206 & 6.805 	& 4.758		& 5.421		& 6.930 \\
index range h	& -9...9	& -5...6		& -5...5	& -17...17	& -6...3 \\
		& -16...16	& -3...6		& -5...5	& -17...17	& -6...5 \\
		& -9...8	& -16...14		& -14...14	& -14...14	& -16...16\\
reflections collected & 14323	& 1821			& 2475		& 6685		& 879 \\
independent \newline reflections & 417 	& 226		& 100		& 703		& 141 \\
R$_\mathrm{int}$ (\%)	& 9.52		& 3.52		& 10.49		& 2.00		& 3.45 \\
RF2 (\%)	& 6.61		& 5.25			& 2.18		& 8.21		& 7.60 \\
RF2w (\%)	& 5.45		& 9.36			& 5.79		& 6.29		& 8.50 \\
RF (\%)		& 3.30		& 3.78			& 2.00		& 5.84		& 4.81 \\
$\chi^2$	& 0.369		& 1.98			& 1.373		& 5.84		& 2.24 \\
no. of free \newline parameters & 19 	& 14		& 14		& 26		& 14 \\
\end{tabularx}
\caption{Experimental details and results of the structure refinement of \cetal\ (\textit{T}\,=\,Ag, Au, Cu, Pd, Pt) as studied at ambient conditions. }
\label{tab:PowderDiff}
\end{center}
\end{table*}

\begin{sidewaystable*}[htbp]
\begin{center}
\scriptsize
\begin{tabularx}{\textwidth}{lXXXXXXXXXXXX}
\vspace{5pt}
Atom	& Wyckoff	& x/a		& y/b		& z/c		& sof 		& u$_{eq}$	& u$_{11}$	& u$_{22}$	& u$_{33}$	& u$_{12}$	& u$_{13}$	& u$_{23}$\\ 
\hline
\multicolumn{13}{l}{\ceagal, space group \textit{Cmcm}, a\,=\,6.2050(12)\,\AA, b\,=\,10.837(2)\,\AA, c\,=\,6.1176(12)\,\AA}  \\
Ce	& 4c		& 0		& 0.24552(6)	& 1/4		& 1.0		& 0.0105(4)	& 0.0097(3)	& 0.0102(3)	& 0.0117(5)	& 0		& 0		& 0 \\
Ag1	& 4c		& 1/2		& 0.38336(9)	& 1/4		& 0.918(8)	& 0.0133(5)	& 0.0125(4)	& 0.0123(5)	& 0.0150(6)	& 0		& 0		& 0 \\
Al1	& 4c		& 1/2		& 0.38336(9)	& 1/4		& 0.082(8)	& 0.0133(5)	& 0.0125(4)	& 0.0123(5)	& 0.0150(6)	& 0		& 0		& 0 \\
Al2	& 4c		& 1/2		& 0.1512(4)	& 1/4		& 1.0		& 0.0132(17)	& 0.0132(17)	& 0.0154(16)	& 0.0111(19)	& 0		& 0		& 0 \\
Al3	& 8e		& 0.2309(4)	& 1/2		& 0		& 0.96(1)	& 0.0140(12)	& 0.0141(10)	& 0.0132(10)	& 0.0148(16)	& 0		& 0		& 0.0002(10) \\
\multicolumn{13}{l}{\ceaual, space group \textit{I4mm}, a\,=\,4.3364(4)\,\AA, c\,=\,10.8501(15)\,\AA } \\
Ce1	& 2a		& 0		& 0		& 0		& 1.0		& 0.0078(7)	& 0.0068(6)	& 0.0068(6)	& 0.0099(7)	& 0		& 0		& 0\\
Au1	& 2a		& 0		& 0		& 0.63564(17)	& 1.0		& 0.0124(6)	& 0.0137(6)	& 0.0137(6)	& 0.0099(5)	& 0		& 0		& 0\\
Al1	& 2a		& 0		& 0		& 0.4064(12)	& 1.0		& 0.008(3)	& 0.003(2)	& 0.003(2)	& 0.017(4)	& 0		& 0		& 0\\
Al2	& 4b		& 0		& 1/2		& 0.2608(7)	& 1.0		& 0.009(3)	& 0.005(3)	& 0.011(3)	& 0.012(3)	& 0		& 0		& 0\\
\multicolumn{13}{l}{\cecual, space group \textit{I4mm}, a\,=\,4.2508(4)\,\AA, c\,=\,10.6436(13)\,\AA }  \\
Ce1	& 2a		& 0		& 0		& 0		& 1.0		& 0.0063(4)	& 0.0058(4)	& 0.0058(4)	& 0.0071(5)	& 0.000		& 0.000		& 0.000\\
Cu1	& 2a		& 0		& 0		& 0.6314(3)	& 0.84(4)	& 0.0084(12)	& 0.0086(13)	& 0.0086(13)	& 0.0080(19)	& 0.000		& 0.000		& 0.000\\
Al1	& 2a		& 0		& 0		& 0.6314(3)	& 0.16(4)	& 0.0084(12)	& 0.0086(13)	& 0.0086(13)	& 0.0080(19)	& 0.000		& 0.000		& 0.000\\
Al2	& 2a		& 0		& 0		& 0.4040(8)	& 1.0		& 0.0092(15)	& 0.0062(17)	& 0.0062(17)	& 0.015(3)	& 0.000		& 0.000		& 0.000\\
Al3	& 4b		& 0		& 1/2		& 0.2489(6)	& 0.94(4)	& 0.0081(13)	& 0.008(3)	& 0.009(3)	& 0.0079(14)	& 0.000		& 0.000		& 0.000\\
\multicolumn{13}{l}{ \ceptal, space group \textit{I4mm}, a\,=\,4.3239(4)\,\AA, c\,=\,10.6670(15)\,\AA } \\
Ce1	& 2a		& 0		& 0		& 0		& 1.0		& 0.0086(14)	& 0.0084(12)	& 0.0084(12)	& 0.0090(19)	& 0.00000	& 0.00000	& 0.00000\\
Pt1	& 2a		& 0		& 0		& 0.6364(3)	& 0.816(10)	& 0.0107(11)	& 0.0127(10)	& 0.0127(10)	& 0.0067(13)	& 0.00000	& 0.00000	& 0.00000\\
Al1	& 2a		& 0		& 0		& 0.6364(3)	& 0.184(10)	& 0.0107(11)	& 0.0127(10)	& 0.0127(10)	& 0.0067(13)	& 0.00000	& 0.00000	& 0.00000\\
Pt2	& 2a		& 0		& 0		& 0.3858(12)	& 0.816(10)	& 0.034(5)	& 0.021(3)	& 0.021(3)	& 0.060(8)	& 0.00000	& 0.00000	& 0.00000\\
Al2	& 2a		& 0		& 0		& 0.3858(12)	& 0.184(10)	& 0.034(5)	& 0.021(3)	& 0.021(3)	& 0.060(8)	& 0.00000	& 0.00000	& 0.00000\\
Al3	& 4b		& 0		& 1/2		& 0.2566(11)	& 1.0		& 0.011(5)	& 0.010(5)	& 0.013(5)	& 0.011(4)	& 0.00000	& 0.00000	& 0.00000\\
\end{tabularx}
\caption{Refined fractional atomic coordinates, site occupations as well as equivalent and anisotropic thermal displacement parameters for \ceagal, \ceaual, \cecual\ and \ceptal.}
\label{tab:S1}
\end{center}
\end{sidewaystable*}

\begin{table*}[htbp]
\begin{center}
\small
\begin{tabularx}{\textwidth}{lXXXXXX} 
\vspace{5pt}
Atom	& Wyckoff	& x/a		& y/b		& z/c		& sof 		& u$_{eq}$	\\
\hline
\multicolumn{7}{l}{ (3a) \cepdal, space group \textit{I4mm}, a\,=\,12.988(1)\,\AA, c\,=\,10.589(1)\,\AA	}\\
Ce1	& 2a		& 0		& 0		& 0		& 1.0		& 0.01266(30)	\\
Ce2	& 8d		& 0.3333(4)	& 0		& 0.0189(9)	& 1.0		& 0.0094(15)	\\
Ce3	& 8c		& 0.3291(4)	& 0.32918	& 0.012(2)	& 1.0		& 0.0129(10)	\\
Pd1	& 2a		& 0		& 0		& 0.649(2)	& 1.0		& 0.0086(17)	\\
Pd2	& 8d		& 0.5		& 0.1653(5)	& 0.875(2)	& 1.0		& 0.0139(25)	\\
Pd3	& 8c		& 0.1696(5)	& 0.16963	& 0.155(2)	& 0.65(2)	& 0.0103(19)	\\
Al1	& 8c		& 0.1696(5)	& 0.16963	& 0.155(2)	& 0.34(2)	& 0.0103(19)	\\
Al2	& 2a		& 0		& 0		& 0.422(3)	& 0.78(7)	& 0.0113	\\
Pd4	& 2a		& 0		& 0		& 0.422(3)	& 0.21(7)	& 0.0113	\\
Al3	& 4b		& 0.5		& 0		& 0.262(6)	& 1.0		& 0.0076(88)	\\
Al4	& 8d		& 0.5		& 0.1687(9)	& 0.102(2)	& 0.80(3)	& 0.0190(57)	\\
Pd5	& 8d		& 0.5		& 0.1687(9)	& 0.102(2)	& 0.19(3)	& 0.0190(57)	\\
Al5	& 8d		& 0.5		& 0.339(3)	& 0.257(5)	& 1.0		& 0.0089(76)	\\
Al6	& 8d		& 0.159(3)	& 0		& 0.271(5)	& 1.0		& 0.0152(89)	\\
Al7	& 8c		& 0.154(2)	& 0.15476	& 0.882(4)	& 1.0		& 0.0241(63)	\\
Al8	& 16e		& 0.347(2)	& 0.168(2)	& 0.260(3)	& 1.0		& 0.0076(51)	\\
\multicolumn{7}{l}{ (BaNiSn$_3$) \cepdal *, space group \textit{I4mm}, a\,=\,4.34396(6)\,\AA, c\,=\,10.5915(2)\,\AA }\\
Ce1		& 2a		& 0		& 0		& 0		& 1.0		& 0.0316(8) \\
Pd1		& 2a		& 0		& 0		& 0.6247(4)	& 1.0		& 0.0410(14) \\
Al1		& 2a		& 0		& 0		& 0.3488(12)	& 1.0		& 0.0114(11) \\
Al2		& 4b		& 0		& 1/2		& 0.2579(8)	& 1.0		& 0.0114(11) \\
\end{tabularx}
\caption{Refined fractional atomic coordinates, site occupations as well as isotropic thermal displacement parameters for two modifications of \cepdal. *results were obtained by the evaluation of X-ray powder diffraction data}
\label{tab:S2}
\end{center}
\end{table*}

\end{document}